# Semantic Technology to Exploit Digital Content Exposed as Linked Data

Riccardo ALBERTONI, Monica DE MARTINO
*CNR-IMATI, Via de Marini 6, Genova, 16149, Italy*
*Tel: +39 010 6575662, Fax: +39 010 6575662,*
Email: albertoni@ge.imati.cnr.it, demartino@ge.imati.cnr.it

**Abstract:** The paper illustrates the research result of the application of semantic technology to ease the use and reuse of digital contents exposed as Linked Data on the web. It focuses on the specific issue of explorative research for the resource selection: a context dependent semantic similarity assessment is proposed in order to compare datasets annotated through terminologies exposed as Linked Data (e.g. habitats, species). Semantic similarity is shown as a building block technology to sift linked data resources. From semantic similarity application, we derived a set of recommendations underlying open issues in scaling the similarity assessment up to the Web of Data.

## 1. Introduction - Issues to be addressed

The paper proposes the research result on semantic technology to ease the use and reuse of digital content exposed as Linked Data resources on the web. In particular, it focuses on the specific issue of supporting explorative research for the resource selection.

Effective sharing and reuse of resources and in particular digital contents (e.g., plain text, documents, images, audio/video, source code) are still desiderata by many scientific and industrial domains, e.g., environmental monitoring and analysis, medicine and bioinformatics, CAD/CAE virtual product modelling and professional multimedia, where the selection of tailored and high-quality content is a necessary condition to provide successful and competitive services. For example, in the domain of environmental data, many data resources are usually obtained through complex acquisition-processing pipelines, which typically involve distinct specialized fields of competency. Oceanographers, biologists, geologists may provide heterogeneous data resources, which are encoded differently in text, tables, images, 2D and 3D digital terrain models. Knowledge management research for the browsing of these contents has to involve issues related to: (i) different user and domain dependant pipelines, (ii) sharing and collaboration between users. For these purposes the use and management of metadata describing digital contents becomes essential.

Semantic Web and in particular the emerging Linked Data [1] provide a promising framework to encode, publish and share complex metadata of resources in these scientific and industrial domains. In particular, the increasing interest for Linked Data is affecting the way information is published, managed, and reused. However, the large part of discussion is still focusing on how to publish and share data rather than how to take advantage of the published Linked Data.

We address the issue about how to exploit Linked Data resources once they have been published providing a step over in their browsing and selection. In particular a context dependent semantic similarity assessment is proposed and applied to compare geographic resources with specific reference to target dataset about habitat in the geographic domain.

Recommendations learnt from this experience in scaling the similarity assessment up to the Web of Data are provided as the final contribution of the paper.

The paper is organized as follows: Section 2 describes the objectives of the paper; Section 3 shows the methodology used; Section 4 provides information about the technology employed and Section 5 shows the application outcomes. Conclusion and recommendations end the paper.

## 2. Objectives

We focus on the issue related to the selection task affecting the browsing activity where users with different skills have to carefully select a set of resources whose metadata is exposed as Linked Data. Semantic similarity will be discussed as example of a set of methods voted at consuming metadata published as linked data. Semantic similarity aims to compare resources identifying those that are conceptually close but not identical, it is proposed as a method supporting the deep comparison among candidates during the resource selection. The methods originally conceived for ontology driven repositories [2] have been extended to the resources published according to Linked Data.

## 3. Methodology Used

We propose a method to analyse digital contents exposed as linked data evaluating their semantic similarity.

The term "semantic similarity" has been used in literature with different meanings. It sometimes refers to ontology alignment, where it enables the matching of distinct ontologies by comparing the names of the classes, attributes, relations, and instances [3]. Semantic similarity can also refer to concept similarity where it assesses the similarity among terms by considering their distinguishing features [4,5,6]; their encoding in lexicographic databases [7,8,9]; and their encoding in conceptual spaces [10].

In this paper, however instance semantic similarity is exploited to support in the comparison of linked data providing different ranking to browse and select them during the search for geographical information.

Different methods to assess instance similarity have been proposed in literature. Some rely on description logics [11]; some have been applied in the context of web services [12]; and some others have been applied to cluster ontology driven metadata [13, 14].

Surprisingly, none of these methods support recognition in the case of those instances, albeit different, have effectively the same informative content: they lack of an explicit formalization of the role of context in the entity comparison, and they fail identifying and measuring if the informative content of one overlaps or is contained in the other. Thus, the similarity results are not easily interpretable in terms of gain and loss the users get adopting a resource in place of another. In this paper, we exploit extension to linked data of the semantic similarity we introduced in [2].

## 4. Technology

From the technological point of view the semantic similarity prototype relies on JENA [15] semantic web framework to retrieve and query RDF Models and to perform some simple RDF reasoning on Linked Data datasets that we consider during the similarity assessment.

It has been applied in the geographic domain within the European project NatureSDIplus (ECP-2007-GEO-317007) to browse a framework of interlinked Knowledge Organization Systems published according to Linked Data. The framework refers to complex domain such as nature conservation. In particular, we have applied the tool to a subset of Habitat types and Species provided by EUNIS database. The two datasets of Habitat types and Species have first been exposed as instances of the standard model

Simple Knowledge Organization System (SKOS) [16], published with D2R server [17] and interlinked according to Linked Data. They are available at the Linked Data server http://linkeddata.ge.imati.cnr.it:2020/.

## 5. Results

In this paragraph we illustrate the setting actions in order to apply the semantic similarity to the species and habitat datasets. Then the results of the assessment are described and discussed.

*5.1 - Identification of a subset of resources to be considered in the similarity assessment*

EUNIS database provides more than 5000 habitats, although from the technological point of view our similarity algorithm has no problem facing the whole EUNIS database, in this paper, we illustrate the result for a subset of published EUNIS Habitats listed in Table 1. Such as a subset corresponds to coastal shingle and part of its subtype.

*Table 1:EUNIS Habitats subset: skos:Concept represents the final part of URIs associated to the SKOS Concepts, skos:title represents the EUNIS Habitat title, skos:relatedMatch summarizes the species associated to each habitat by the interlinking procedure*

| skos:Concept | skos:title | skos:relatedMatch |
|---|---|---|
| B2 | Coastal shingle | None. |
| B2.1 | Shingle beach driftlines | [Atriplex], [Cakile maritima], [Glaucium flavum], [Euphorbia paralias], [Euphorbia peplis], [Eryngium maritimum], [Matthiola sinuata], [Matthiola tricuspidata], [Mertensia maritima], [Polygonum], [Salsola kali] |
| B2.11 | Boreo-arctic gravel beach annual communities | [Atriplex longipes], [Atriplex glabriuscula], [Cakile edentula], [Mertensia maritima], [Polygonum norvegicum], [Polygonum oxyspermum ssp. raii] |
| B2.12 | Atlantic and Baltic shingle beach drift lines | [Atriplex] [Atriplex glabriuscula), [Cakile maritima ssp. maritima], [Cakile maritima ssp. baltica], [Euphorbia peplis], [Glaucium flavum], [Mertensia maritima], [Polygonum], [Salsola kali] |
| B2.13 | Gravel beach communities of the mediterranean region | [Atriplex], [Cakile maritima ssp. aegyptiaca], [Cakile maritima ssp. euxina], [Enarthrocarpus arcuatus], [Eryngium maritimum] [Euphorbia peplis], [Euphorbia paralias], [Glaucium flavum], [Matthiola sinuata], [Matthiola tricuspidata], [Salsola kali], [Polygonum] |
| B2.14 | Biocenosis of slowly drying wracks | None |
| B2.3 | Upper shingle beaches with open vegetation | [Crambe maritima], [Honkenya peploides], [Lathyrus japonicus] |
| B2.31 | Baltic sea kale communities | [Angelica archangelica ssp. litoralis], [Atriplex], [Beta vulgaris ssp. maritima], [Crambe maritima], [Elymus arenarius], [Elymus repens], [Euphorbia palustris], [Geranium robertiana ssp. rubricaule], [Glaucium flavum], [Honkenya peploides], [Isatis tinctoria], [Leymus arenarius], [Ligusticum scoticum], [Mertensia maritima], [Silene vulgaris ssp. maritima] [Silene uniflora], [Tripleurospermum maritimum], [Valeriana salina] |
| B2.32 | Channel sea kale communities | [Crambe maritima], [Honkenya peploides], [Lathyrus japonicus] |
| B2.33 | Atlantic sea kale communities | [Crambe maritima], [Crithmum maritimum] [Beta vulgaris ssp. maritima], [Galium aparine], [Glaucium flavum], [Matricaria Maritima], [Rumex crispus], [Solanum dulcamara var. maritima], [Sonchus oleraceus] |
| B2.34 | Gravelly beach and | [Ammophiletea], [Agropyro juncei-Sporoboletum pungentis] |

| | shingle pioneer communities | [Medicagini marinae-Triplachnion nitensis] |

*5.2 - Identification and formalization of two different contexts*

Following the formalism we have introduced in [2], we have defined two distinct contexts:
1. Context 1: to compare (i) habitats according to the species that they host (hereafter it is also referred as "habitat species-based similarity") and (ii) the species according to habitats they live into.
   PREFIX skos: <http://www.w3.org/2004/02/skos/core#>
   [skos:Concept]->{{},{(skos:relatedMatch, Inter)}}
2. Context 2: to compare habitats instances with respect to their position in the taxonomy hierarchy (hereafter, it is also referred as "taxonomy-based similarity").
   PREFIX skos: <http://www.w3.org/2004/02/skos/core#>
   [skos:Concept]->{ {},{(skos:broader, Inter)}}

The first line of both contexts specifies a XML name space, which allows to use "skos:" as abbreviation for http://www.w3.org/2004/02/skos/core#. In the first context, two instances of skos:Concept are compared with respect to the skos:Concept related by skos:relatedMatch. The more two instances of skos:Concept match with common related instances of skos:Concept the more they are similar. Considered how the skos:relatedMatch have been added to species and habitats in the described interlinking we can exploit such a context to compare habitats according to the species that they host and the species according to habitats they pertain to.

In the second context, two instances of skos:Concept are compared with respect to their broader skos:Concept, namely the skos:Concept related by skos:broader. Species and habitats are organized in semantically meaningful taxonomies. The position in the taxonomy is often exploited to determine the semantic similarity between the entities [13]. For the purpose of this similarity assessment, we consider the skos:broader relation transitive and reflexive. The skos:broader transitive and reflexive closure is materialized adding the following rule to the JENA reasoner

(?x skos:broader ?y) (?y skos:broader ?z)-> (?x skos:broader ?z)
(?y skos:broader ?z)-> (?y skos:broader ?y)

As a result of the reasoning induced by this rule, ancestors of each skos:Concept instance are added to each instance as skos:broader, thus the comparison with respect to position in the taxonomy can be performed by comparing the instances directly related by skos:broader.

*5.3 - Experiment result and discussion*

Semantic similarity assessments among habitats with respect to the two contexts previously formalized are illustrated in Figure 1: Figure 1 (a) shows the result related to the context 1 and Figure 1 (b) shows the result related to the context 2. Each column i and each row j of the matrix represents a habitat. The grey level of the pixel (i,j) represents the similarity value (SIM(i,j)) between the two habitats located at row j and column i: the darker is the colour, the more similar are the two habitats.

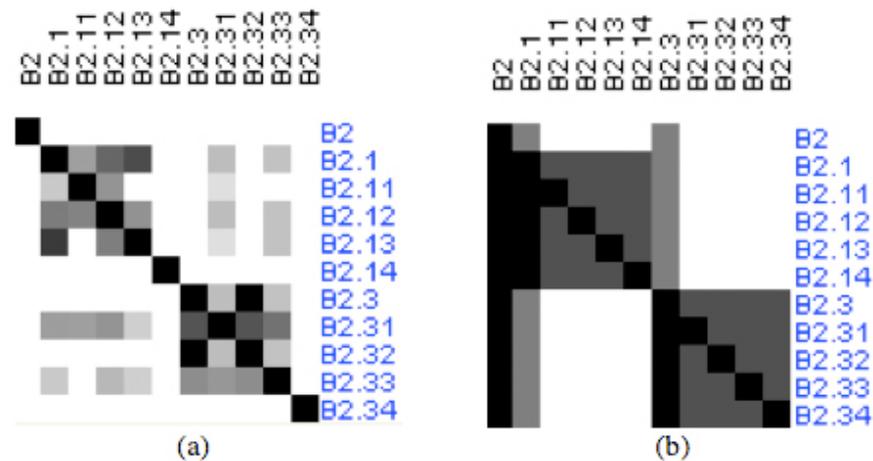

*Figure 1: Two similarity matrices of similarity comparing the subset of habitats of Table1 with respect to (a) context 1 and (b) context 2*

Even if the similarity is calculated on a subset of habitats, the similarity results provide elements to draw some interesting considerations:
- The similarity assessment works out differently according to the context we apply: matrices in Figure 1(a) and Figure 1(b) look completely different. For example, according to the taxonomy-based similarity in Figure 1(b), B2.1 (Shingle beach drift lines) shares nothing with B2.31 (Baltic sea kale communities) and B2.33 (Atlantic sea kale communities), the contrary, they seem to have something in common when compared on species-base similarity as Figure 1(a)
- The similarity matrices are asymmetric, asymmetry can be exploited to explore in depth the relations among habitats. For example, in Figure 1(a), SIM(B2.31, B2.1) > SIM(B2.1, B2.31), and that means B2.31 (Baltic sea kale communities) shares more of its species with B2.1 (Shingle beach driftlines) than vice versa. Actually, considering Table 1, B2.31 shares with B2.1 3 out of 18 species, whilst B2.1 shares with B2.31 3 out of 11 species
- The Figure 1(b) gives proof of the informativeness associated to containment highlighting. As a consequence of the containment highlighted by asymmetry, the taxonomic structure of habitats is actually shown by the columns of black dots in Figure 1(b). For example, observing Figure 1(b), we can note that B.2 is contained in all the other habitats because its column has only black pixels. B.2 is related to the other habitats by skos:broader, because it is ancestor of all the other habitats. Thus SIM(B.2, X)=1 for any X, and B.2 column is made of black dots. Similarly, we can notice in Figure 1(b) B2.1 is the ancestor of B2.11, B2.12, B2.13 and B.14, and that B2.3 is ancestor of B2.31, B2.32, B2.33 and B2.34.

More in general, asymmetric similarity assessment enables new way to browse and query resources exposed as linked data. It is extremely powerful since the highlighting of containment provides sound interpretation of the similarity results. For example, the similarity assessed according to context 1 can be exploited to rank habitats that share species. Given two habitats X, Y, the similarity results can be interpreted as follows:
- if SIM(X,Y)=1 and SIM(Y,X)=1 then Y contains the same species of X
- if SIM(X,Y)=1 and SIM(Y,X)<1 then Y contains the species of X but the vice versa is not true
- SIM(X,Y) is proportional to the percentage of species in X that are contained in Y out of the overall species of X.

Since, we have interlinked habitats to species and vice versa, we can also compare species with respect to habitats they are related to. For example, we have considered the list of species, which are interlinked to the habitats in Table 1, and we have applied the semantic similarity according to the context 1. The results are illustrated in Figure 2.

The comparison of species with respect to habitats considers the list of species in Table 1. However a species might be related to habitats that are not listed in the table, that habitats even if not listed in Table 1 are considered in the similarity assessment.

Even in this case the asymmetry of semantic similarity, which is designed to exploit the containment, is useful to interpret the similarity results.

Given two species X, Y, the interpretation of similarity results is:
1. if SIM(X,Y)=1 and SIM(Y,X)=1 then Y contains the same habitats of X
2. if SIM(X,Y)=1 and SIM(Y,X)<1 then Y contains the habitats of X but the vice versa is not true
3. SIM(X,Y) is proportional to the percentage of habitats in X that are contained in Y out of the overall habitats of X.

*Figure 2: Similarity matrix resulting comparing species in Table 1 with respect to Habitats they are related to (according to context 1)*

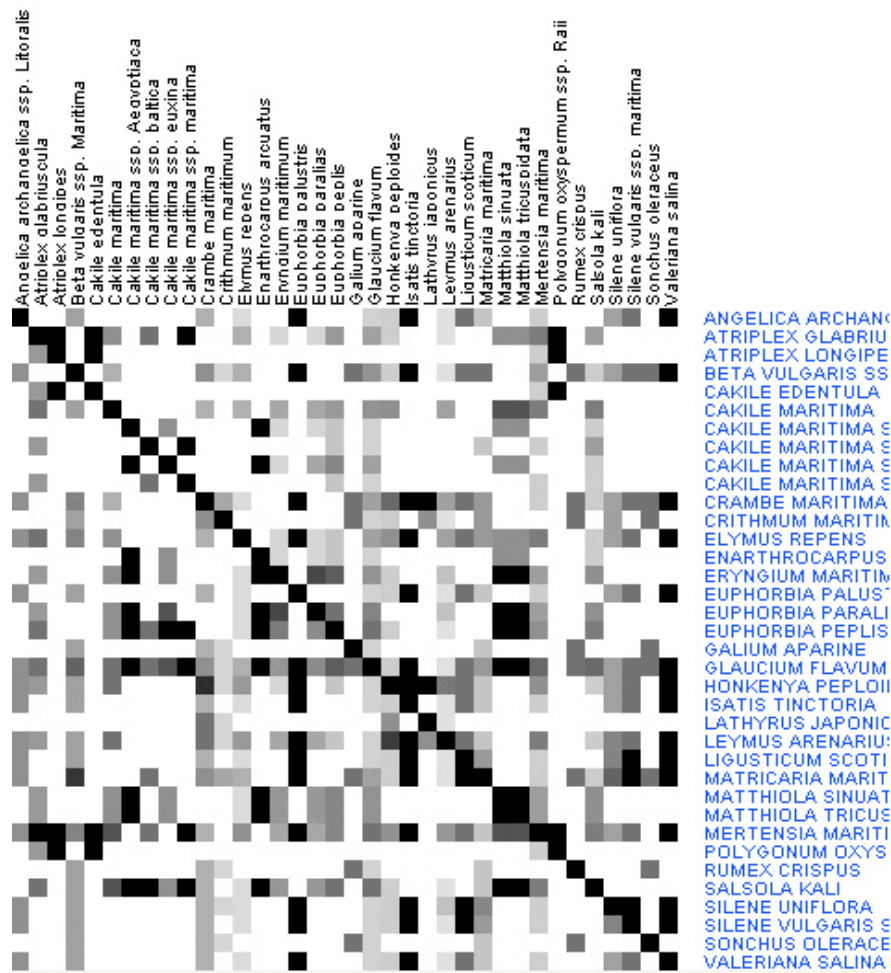

Interesting considerations also arise by observing the matrix in Figure 2:
- some columns such as those pertaining to "valeriana salina", "cakile edentula" results equal to zero (i.e., white dots) or one (i.e., black dots). That is a consequence of the containment highlighting. These species characterizes only few habitats: for example, in the EUNIS database "valeriana salina" is a species associated to "Baltic [Crambe maritima] communities" only. Thus, the similarity between "valeriana salina" and a

species X results (a) "1" whenever X is related to "Baltic [Crambe maritima] communities", (b) "0" otherwise
- containment between species with respect to habitats is highlighted. For example the pixel corresponding to SIM("Atriplex longipes", "Atriplex glabriuscula") is black then the similarity is equal to 1 and the pixel corresponding to SIM("Atriplex glabriuscula","Atriplex longipes") is grey then the similarity is less than 1. That means that the habitats in which "Atriplex longipes" lives are a sub set of habitats in which "Atriplex glabriuscula" also lives. Containment makes explicit this kind of observations, which are potentially very informative and can be a source of further investigation in complex case use cases.

As overall considerations about the application of the proposed context dependent asymmetric similarity, we notice:
- the different contexts provide means to rank and browse habitats according to specific features paving the way for defining a browsing customized with respect to specific user's views;
- the more datasets are interlinked, the more new contexts can be defined enabling new way to sift data;
- the asymmetric similarity assessment emphasises the containment between compared elements, which is a powerful tool interpreting the similarity results. It makes explicit patterns that are bases for a deeper comprehension of data and source for further investigation among the underlining nature of data.

## Conclusions and Summary Recommendations

In the paper we have illustrated the application of semantic technology to select digital contents whose metadata is exposed as Linked Data on the web. As an example, we have shown how our semantic similarity supports in sifting SKOS terminologies annotating geographical datasets. The example compares resources in complex domain such as nature conservation and in particular related to habitats and species. We can consider it as a first technique to analyse Linked Data resources. Techniques such as semantic similarity can be combined with other search tools (e.g., facet search) in order to get extremely powerful exploratory search.

Unfortunately, from our experience we have learned that there are still many issues that have to be addressed to scale the instance similarity up to the web of data. In particular, as recommendation for future research we identify the following issues:
- non-authoritative metadata (i.e. metadata published by actors who are neither the resource producers nor the owners) can be investigated considering how synergies with semantic web indexes (e.g., Sindice) can be used to retrieve non authoritative features
- heterogeneous metadata, (i.e. metadata provided according to different, sometimes interlinked, more often overlapping metadata vocabularies) can be addressed deploying entity level consolidation using both explicit metadata statements and mining implicit equivalences through co-occurring resources annotations
- non-consistently identified metadata (e.g., metadata occurring when the same resource has different identifiers in distinct metadata sets) could be eased deploying reasoning techniques to be applied to web datasets
- efficiency and computational issue: in the long term an accurate similarity assessment might result computationally prohibitive as soon as the number of resources discovered and features considered increase. It can deploy strategies to speed up the assessment of semantic similarity, in particular, solutions based on the cashing of intermediate comparisons and techniques to prune the comparisons according to a specified application context might resolve the less severe cases. Moreover, algorithms for efficient parallelization can be studied, e.g., using the Map Reduce cluster-computing paradigm.

## Acknowledgements

This activity has been partially supported within the EU project NatureSDIplus (ECP-2007-GEO-317007) of the eContentPlus program.